\documentclass{article}

\usepackage{arxiv}

\usepackage[utf8]{inputenc} 
\usepackage[T1]{fontenc}    
\usepackage{hyperref}       
\usepackage{url}            
\usepackage{booktabs}       
\usepackage{amsfonts}       
\usepackage{nicefrac}       
\usepackage{microtype}      
\usepackage{lipsum}		
\usepackage{graphicx}
\usepackage{natbib}
\usepackage{doi}
\usepackage{threeparttable} 
\usepackage{amsmath,bm} 
\usepackage{bbm} 
\usepackage{subcaption} 
\usepackage{derivative} 

\title{Predictive modeling for limited distributed targets}

\author{ \href{https://orcid.org/0000-0000-0000-0000}{\includegraphics[scale=0.06]{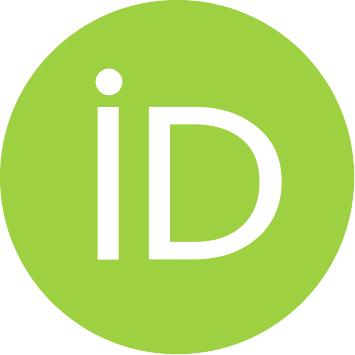}\hspace{1mm}Björn Bokelmann} \\
	Chair of information systems\\
	Humboldt University Berlin\\
	Unter den Linden 6\\
    10099 Berlin\\
	\texttt{bokelmab@hu-berlin.de} \\
	\And
	\href{https://orcid.org/0000-0000-0000-0000}{\includegraphics[scale=0.06]{orcid.pdf}\hspace{1mm}Stefan Lessmann} \\
	Chair of information systems\\
	Humboldt University Berlin\\
	Unter den Linden 6\\
    10099 Berlin\\
}

\renewcommand{\shorttitle}{Binarized target modeling}

\hypersetup{
pdftitle={Binarized target modeling},
pdfsubject={q-bio.NC, q-bio.QM},
pdfauthor={Björn~Bokelmann, Stefan~Lessmann},
pdfkeywords={First keyword, Second keyword, More},
}

\begin{document}
\maketitle

\begin{abstract}
Many forecasting applications have a limited distributed target variable, which is zero for most observations and positive for the remaining observations. In the econometrics literature, there is much research about statistical model building for limited distributed target variables. Especially, there are two component model approaches, where one model is build for the probability of the target to be positive and one model for the actual value of the target, given that it is positive. However, the econometric literature focuses on effect estimation and does not provide theory for predictive modeling. Nevertheless, some concepts like the two component model approach and Heckmann's sample selection correction also appear in the predictive modeling literature, without a sound theoretical foundation. In this paper, we theoretically analyze predictive modeling for limited dependent variables and derive best practices. By analyzing various real-world data sets, we also use the derived theoretical results to explain which predictive modeling approach works best on which application.   
\end{abstract}


\section{Introduction}\label{introduction}
In the last decade, predictive models have gained importance in a very wide range of domains. In the twentieth century, traditional statistical models, like linear regression models, logistic regression models and other generalized linear models were most frequently applied for forecasting applications. These models require an explicit specification, by an equation including free parameters, of the relationship between the target to predict and the features. Although useful, these traditional statistical models were limited in their predictive performance, as they can not describe complex, non-linear relationships between features and target. This limitation of the traditional models, combined with a growing amount of large data sets and a rise of computational power were the reasons for a gain in importance of machine learning methods, starting around the beginning of the twenty first century. The novel forecasting methods do not require an explicit specification of the relationship between features and target and are able to represent complex, non-linear relationships.

Our research does not focus on the choice of forecasting model, but rather focuses on the statistical properties of the target variable to predict, which affect whatever model is applied. We consider forecasting a target variable $y$, with the following distribution properties: $y$ is zero with a positive probability and if it is not zero, it has a distribution of positive values. We denote this kind of distribution a \textit{limited distribution}. There are many forecasting applications were the target variable has such a limited distribution. In medicine, most of the patients are not readmitted ($y=0$) after being discharged from the hospital. But a considerable share of those being readmitted, get readmitted multiple times ($y>0$) \citep{bardhan2015predictive}. In the same way, many customers do not consume certain products, like tabaco, at all ($y=0$) but those who consume them do this frequently ($y>0$) \citep{badicu2020predicting}. In online marketing, a big share of market visitors does not buy anything ($y=0$), but customers who buy something spend a positive amount of money ($y>0$) \citep{gubela2020response}. When a bank grants credits, most of the borrowers do not default ($y=0$), but some do and this causes a certain monetary loss ($y>0$) to the bank \citep{yao2017enhancing}. The list of forecasting applications where the target has such a limited distribution could be further extended. 

Building statistical models for forecasting and inference customized to such limited distribution of $y$ has already been considered in various econometric studies. The first study proposing a statistical modeling strategy for such limited distributions was conducted by \citet{tobin1958estimation}. He found that for limited target distributions, a linear regression model would fail the assumption of linearity as well as the assumption of normally distributed error terms. To overcome this problem, he proposed the \textit{Tobin model}, which is a two component model, consisting of a classification model for the probability of $y>0$ and a regression model for $y$, given $y>0$. \citet{cragg1971some} extended the Tobin model by relaxing assumptions about the relationship between features and targets in the classification and the regression model. The resulting models are called \textit{hurdle models}. In this paper, we summarise any approach which uses one model for the probability of $y>0$ and another model for the value of $y$, given $y>0$, under the name \textit{two component models}. Another important aspect when building statistical models for limited target variables comes from \citet{heckman1979sample} and concerns the problem of sample selection when building a regression model for $y$, given $y>0$. He analyzed the resulting bias when a model is naively fitted only on the observations with $y>0$ and suggested a method to overcome this bias. 

So, the econometric literature has extensively studied statistical modeling for limited target variables. But it is unclear whether their insights can be transferred to the case of \textit{predictive modeling} with \textit{modern machine learning methods}. This is because the econometric literature concerns traditional statistical models, with their assumptions about the form of relationship between features and target and assumptions about the target distribution. Violations of these assumptions for limited target variables led to the use of two component models. But for flexible forecasting methods like neural networks, random forest, etc. these assumptions do not apply. Nevertheless, there is a range of studies applying variants of the two component models for forecasting, partly using machine learning methods. Examples are \citet{zadrozny2001learning} who apply a two component model approach to predict expected donation amounts and who apply Heckman's methods of sample selection bias prevention, \citet{haupt2022targeting} and \citet{baier2022profit} who apply two component models for predicting causal effects in online marketing. \citet{baier2022profit} even apply Heckman's correction. Furthermore, \citet{yao2017enhancing, loterman2012benchmarking} apply two component models to predict the monetary risk of granted credits. Still, there lacks theoretical motivation for such approaches.

Our research question is to analyze the statistical properties of limited dependent variables and the consequences for best practice in predictive modeling. We show that building separate models for the probability of $y>0$ and the value of $y$, given $y>0$, can be beneficial due to an increase in the signal-to-noise ratio, when compared with a single model for the target $y$. After theoretically proving potential benefits of the two component model approach, we also examine potential problems. The two component models in are not optimized to predict the original target $y$. Accordingly, their combination as a product is not necessarily an accurate model for $y$. More precisely, we demonstrate that scale (variation in the predictions of the sub-models) and location (mean predictions of the sub-models) are not necessarily ideal and show how this problem can be fixed. We call this problem the \textit{scaling and location problem}. We also theoretically analyse the problem of sample selection bias when building a component model for the amount of $y$ on the sub-sample of observations with $y>0$. We conclude that this selection bias does not pose a severe problem and that a Heckman correction, as applied in some previous studies for predicting limited dependent variables, lacks theoretical justification.

To demonstrate the practical value of our theoretical research results, we consider three real-world examples of predictive modeling for limited dependent variables. One application is the decision of whom to contact for a charity donation request. The other two applications come from online marketing.

\section{What makes a good predictive model?}
\subsection{Predictive models and error metrics}
Our paper considers the case of a prediction task. Thereby, a certain target variable $y$ should be predicted, given features $x$. A predictive model $\hat{f}(x)$ is build on training data, such that it ideally predicts $y$ on separate test data. In this section, we reflect on the metrics to evaluate the model predictions and motivate our choice for the following theoretical analysis.  

The most famous metric for regression model evaluation is the mean squared error. It measures the mean of the squared differences between the model predictions and the prediction target $y$ on the test set. Thereby, the calculated metric on the test set can be seen as an estimate for the theoretical model performance, given by
\begin{align*}
MSE(\hat{f}(x),y)=E[(y-\hat{f}(x))^2].    
\end{align*} 

To better understand, what the $MSE$ measures, it is useful to derive the following expression
\begin{align*}
MSE(\hat{f}(x),y)&=E[(y-E[y]+E[\hat{f}(x)]-\hat{f}(x)+E[y]-E[\hat{f}(x)])^2] \\
&=E[(y-E[y])^2]+E[(E[\hat{f}(x)]-\hat{f}(x))^2]+(E[y]-E[\hat{f}(x)])^2+2E[(y-E[y])(E[\hat{f}(x)]-\hat{f}(x))]\\
&=Var[y]+Var[\hat{f}(x)]-2Cov(\hat{f}(x),y)+(E[y]-E[\hat{f}(x)])^2\\
&=Var[y]+Var[\hat{f}(x)]-2Cor(\hat{f}(x),y)\sqrt{Var[\hat{f}(x)]Var[y]}+(E[y]-E[\hat{f}(x)])^2
\end{align*} First, note, that the higher a models correlation with the target variable, the lower its $MSE$. Now, consider two models $\hat{f}_{1}(x)$ and $\hat{f}_{2}(x)$, which achieve the same correlation with the target variable $y$. If their $MSE$ values differ, this could only be due to differences in $Var[\hat{f}_{i}(x)]$ or $E[\hat{f}_{i}(x)]$. In other words, performance differences in terms of $MSE$ for models having the same correlation with the target variables, can only be due to differences in their respective scaling ($Var[\hat{f}_{i}(x)]$) or location ($E[\hat{f}_{i}(x)]$). This means, the $MSE$ evaluates the correlation of a model with its target and scaling and location of the model. Note, that scaling and location of the model are optimal according to the $MSE$, if $Var[\hat{f}(x)]=Cor(\hat{f}(x),y)^2Var[y]$ and $E[\hat{f}(x)]=E[Y]$.

In our theoretical analysis, we chose to consider the correlation between model predictions and the model target as the performance measure of primary interest, instead of the $MSE$. We did this for two reasons: First, in many applications, scaling and location are irrelevant. If, for example, the goal is to rank instances, scaling and location do not matter. Note, that we do not base our theoretical analysis on ranking metrics like the lift curve, because it is very hard to analyse them statistically. However, the correlation between predictions and target is clearly strongly related to lift. Second, it is worth to separate the evaluation of scale and location from the evaluation of the correlation, because the sub-optimal scaling and location problem of two component models requires distinct analysis. 

\subsection{Observable and theoretical targets}
In this section, we introduce two concepts, which are fundamental for the theoretical analysis of predictive models: the \textit{conditional expected value} $\zeta_{x}:=E[y|x]$, which describes the variation in $y$, explainable by the features $x$ and the \textit{noise} $\varepsilon$, which describes the remaining variation in $y$, being unrelated to the features. It holds
\begin{align*}
y=\zeta_{x}+\varepsilon.
\end{align*}  

Having decomposed $y$ in conditional expectation and noise, we can derive the following equations for the performance of a model $\hat{f}(x)$, measured by the correlation with $y$:
\begin{align}
Cor(\hat{f}(x),y)&=\frac{Cov(\hat{f}(x),y)}{\sqrt{Var[\hat{f}(x)]}\sqrt{Var[y]}}\nonumber\\
&=\frac{Cov(\hat{f}(x),\zeta_{x})+Cov(Cov(\hat{f}(x),\varepsilon)}{\sqrt{Var[\hat{f}(x)]}\sqrt{Var[\zeta_{x}]+Var[\varepsilon]}}\nonumber\\
&=\frac{Cov(\hat{f}(x),\zeta_{x})}{\sqrt{Var[\hat{f}(x)]}\sqrt{Var[\zeta_{x}]+Var[\varepsilon]}}\nonumber\\
&=Cor(\hat{f}(x),\zeta_{x})\sqrt{\frac{Var[\zeta_{x}]}{Var[\zeta_{x}]+Var[\varepsilon]}}\label{eq_th_cor}
\end{align} Accordingly, the correlation of $\hat{f}(x)$ with the target $y$ depends on the correlation of $\hat{f}(x)$ with the conditional expected value $\zeta_{x}$ as well as on the ratio between the variance in $\zeta_{x}$ and the variance in $y$. Note, that the model $\hat{f}(x)$ only affects the first factor. Thus $Cor(\hat{f}(x),\zeta_{x})$ is what really contains the performance of $\hat{f}(x)$. The second factor only contains statistical properties of the target variable.   

Given this decomposition of the correlation performance measure, we can motivate the distinction in \textit{observable and theoretical targets}. We call $y$ the observable target, because it is observable and we can use it to calculate a performance metric $Cor(\hat{f}(x),y)$ on the test set. In contrast, we call $\zeta_{x}$ the theoretical target, because $Cor(\hat{f}(x),\zeta_{x})$ is what really measures the performance of $\hat{f}(x)$. But as we can not observe the theoretical target, we would need to calculate the correlation with the observable target in practice. For the theoretical analysis, however, it is useful to examine the correlation with the theoretical target. This is why, in the following, we always analyse model performance with regard to the theoretical target. 

\subsection{The signal-to-noise ratio}\label{sec_sig_noise}
After having discussed the problem of model evaluation, we now reflect on what properties of the data allow building accurate predictive models. Thereby, in contrast to most of the literature on predictive modeling, we do not focus on the choice of model (e.g. random forest vs. neural networks) but rather on the statistical properties inherent in the data, which would affect whatever model is applied. The most important statistical measure for our theoretical analysis is the \textit{signal-to-noise ratio}, which measures the relative impact of explainable and unexplainable variation in $y$
\begin{align*}
SNR_{y}:=\frac{Var[\zeta_{x}]}{Var[\varepsilon]}.    
\end{align*}

The signal-to-noise ratio has a strong impact on predictive performance. From equation \eqref{eq_th_cor}, we can derive
\begin{align*}
Cor(\hat{f}(x),y)=Cor(\hat{f}(x),\zeta_{x})\sqrt{\frac{SNR_{y}}{SNR_{y}+1}}.    
\end{align*} So, the signal-to-noise factor is a limiting factor on the correlation with the observable target. Even if the model is perfect (e.g. $\hat{f}(x)=\zeta_{x}$) the model performance measured on the test set would still be low, if the signal-to-noise ratio $SNR_{y}$ is low. But the signal-to-noise ratio also has another effect. The lower the signal-to-noise ratio on the training data, the harder it becomes for a model building algorithm to extract the signal $\zeta_{x}$. In consequence, for any predictive model, the lower the signal-to-noise ratio, the lower the correlation $Cor(\hat{f}(x),\zeta_{x})$ will tend to be.

To round off the discussion of the statistical properties, we finally consider the training sample size. It is common knowledge, that the performance of a predictive model increases with increasing training sample size. But following on the previous points, we can further specify this statement. An increase in the training sample size will not enable any model to perform better than $\zeta_{x}$, so the signal-to-noise ratio will still be a limiting factor on the model performance measured with the observable target. However, an increase in the training sample will enable the model building algorithms to better extract the signal $\zeta_{x}$, because noise "averages out" with an increasing number of observations. Accordingly, with increasing training set size, predictive models $\hat{f}(x)$ would achieve a higher correlation with $\zeta_{x}$.

In other studies about predictive models, the signal-to-noise ratio and its impact on predictive models is rarely discussed. Why does this discussion matter for our research question? In contrast to other prediction problems, where the target of predictive models is fixed, limited dependent variables grant some scope for the definition of prediction targets. We could either build a single model for the target $y$, or we could build a two component model, consisting of a model for the binary target $\mathbbm{I}_{y>0}$ and a model for the target $y$, on the observations with $y>0$. The important aspect here is that the signal-to-noise ratio for the single model approach will be different from the signal-to-noise ratios of the targets for the two component model approach. It is exactly this difference in the signal-to-noise ratio which motivates the two component model approach compared to the single model approach. In the next section, we will elaborate why using the two model approach often yields a higher signal-to-noise ratio for the two component models and can thereby lead to improved predictive performance of the two component model approach compared to the single model approach.      

\section{Predictive models for limited distribution targets}

\subsection{Definition of prediction targets}\label{sec_def_target}
In the following, we want to derive a representation for limited dependent variables 
\begin{align*}
y=c\cdot a,    
\end{align*} as the product of a binary variable $c$, which takes the values 1 and 0 and a continuous variable $a>0$. Clearly, such a representation is valid, once we define $a:=y$, when $y>0$ and $c:=\mathbbm{I}_{y>0}$. For the following analysis, it is necessary to decompose $a$ and $c$ in their respective conditional expected value and their respective noise. We denote this decomposition as 
\begin{align*}
c&=p_{x}+\varepsilon_{p}\\
a&=\mu_{x}+\varepsilon_{\mu}.
\end{align*} Thereby, $p_{x}$ is the conditional expected value of $c$ over the whole population and $\mu_{x}$ is the conditional expected value in the sample with $c=1$.  

Next, we need to derive a decomposition in the expected value and the noise for the target variable $y$, with regard to the components above.´It holds
\begin{align*}
E[y|x]&=P[c=1|x]E[y|c=1,x]+P[c=0|x]E[y|c=0,x]\\
&=P[c=1|x]E[a|c=1,x]\\
&=p_{x}\mu_{x}
\end{align*} Because
\begin{align*}
y&=(p_{x}+\varepsilon_{p})(\mu_{x}+\varepsilon_{\mu})\\
&=p_{x}\mu_{x}+p_{x}\varepsilon_{\mu}+\varepsilon_{p}\mu_{x}+\varepsilon_{p}\varepsilon_{\mu},
\end{align*} $y$ has the decomposition
\begin{align}
y&=p_{x}\mu_{x}+\varepsilon \label{eq_dec_y} \text{ with }\\
\varepsilon&=p_{x}\varepsilon_{\mu}+\varepsilon_{p}\mu_{x}+\varepsilon_{p}\varepsilon_{\mu}.\nonumber
\end{align}

\subsection{Signal and noise in limited dependent variables}
In the following, we examine signal and noise of limited dependent variables. Being a product of two variables $a$ and $c$, they naturally obtain some of the signal contained in both variables, but also some of the noise. In the following, we analyse how signal and noise of the component target enter the composite target $y=ac$. In Appendix \ref{app_sig_noise}, we derived the equation
\begin{align*}
SNR_{y}=\alpha\cdot SNR_{a}+(1-\alpha)SNR_{c}+\Psi,    
\end{align*} where $\Psi$ is a component, depending on the relationship between $a$ and $c$ and $\alpha\in [0,1]$. What do we obtain from this equation? Under mild conditions, which can be seen in Appendix \ref{app_sig_noise}, $\Psi$ is negative and thus $SNR_{y}$ is less than a weighted sum of $SNR_{a}$ and $SNR_{c}$. So, in this case, at least one of the component targets has a higher signal-to-noise ratio than $y$. It is well possible, that both component targets have a higher signal-to-noise ratio than $y$. Even if the relationship between $a$ and $c$ is such that $\Psi>0$, it is well possible that one of the component targets has a higher signal-to-noise ratio than $y$.

So, given this analysis about the signal-to-noise ratio, it is very likely that separating a limited dependent variable $y=ac$ in the components $a$ and $c$, yields a higher signal-to-noise ratio of at least one of these components, compared to $y$. In consequence, we would expect the model for the component target with this favorable signal-to-noise ratio to achieve a higher correlation with its component target, than the single model $\hat{\zeta}(x)$ achieves for the target $p_{x}\mu_{x}$. But this is only one half of the picture. The fact that one or both components achieve a good performance for their respective component targets does not yet imply, that their combination is a good predictive model for the composite target $p_{x}\mu_{x}$. The missing half of the picture will be delivered in the next section.

\subsection{Linear relationship between the composite target and the component targets}\label{sec_comp_models}
In this section, we will examine what happens if we build predictive models for the component targets $a$ and $c$, instead of for the target $ac$. We already know that a low signal-to-noise ratio in $y=ac$ could be a reason to try model building for the component targets. However, we do not know, when high predictive accuracy of a model for $a$ or for $c$ translates into high predictive accuracy for $ac$. This question will be answered in this section. 

Before diving into the hard theory, we start with an intuition: The theoretical target to predict is a product $p_{x}\mu_{x}$ of the theoretical component targets. The variance in $p_{x}\mu_{x}$ is driven by the variance  in $p_{x}$, the variance in $\mu_{x}$ and some variance due to the interaction between $p_{x}$ and $\mu_{x}$. Now, consider the case that either $p_{x}$ or $\mu_{x}$ are the main driver of the variance in $p_{x}\mu_{x}$ and the interaction between both theoretical targets has a rather small effect. Then it seems plausible, that a model which predicts $p_{x}$ or $\mu_{x}$ (whatever is the main driver) well, will also be a good model for $p_{x}\mu_{x}$. So, if one of the component targets has high influence on $p_{x}\mu_{x}$ and a high signal-to-noise ratio compared to $SNR_{y}$, it is likely that a model build for this target could predict $p_{x}\mu_{x}$ better than a single model trained for the target $ac$. By combining this valuable component target model with the model for the other component target, the predictive accuracy for $p_{x}\mu_{x}$ could possibly be even more increased.

In the following, we will describe this mathematically. A model trained on the observations $(c_{i},x_{i})$, will be denoted $\hat{p}(x)$ in the following. Its theoretical target is $p_{x}$. A model trained on the observations $(a_{i},x_{i})$, for which $c_{i}=1$, will be denoted $\hat{\mu}(x)$ in the following. Its theoretical target is $\mu_{x}$. We start by analyzing which predictive performance these component models could individually achieve for the composite theoretical target $p_{x}\mu_{x}$. They are trained to achieve as high a correlation $Cor(\hat{p}(x),p_{x})$ respectively $Cor(\hat{\mu}(x),\mu_{x})$ as possible. How does this performance for their individual theoretical component targets translate into a performance for the composite target $p_{x}\mu_{x}$? As we show in Appendix \ref{app_class_model}, they would respectively achieve a performance of
\begin{align*}
Cor(\hat{p}(x),p_{x}\mu_{x})&\approx Cor(\hat{p}(x),p_{x})Cor(p_{x},p_{x}\mu_{x})\\
Cor(\hat{\mu}(x),p_{x}\mu_{x})&\approx Cor(\hat{\mu}(x),\mu_{x})Cor(\mu_{x},p_{x}\mu_{x}).
\end{align*}   

So, if $Cor(p_{x},p_{x}\mu_{x})$ is high (e.g. close to 1) and the signal-to-noise ratio of $c$ is much higher than the signal-to-noise-ratio of $y$, it is very likely that $\hat{p}(x)$ predicts $p_{x}\mu_{x}$ better than a single model $\hat{\zeta}(x)$ trained for the observable target $y=ac$ would do. The same holds true for $\hat{\mu}(x)$, if $Cor(\mu_{x},p_{x}\mu_{x})$ is high and the signal-to-noise ratio of $a$ is higher than the signal-to-noise ratio of $y$. With regard to our intuitive reflection on the component models performance, we can see that the correlation between a theoretical component target and the theoretical composite target is the mathematical expression for the "influence" of the respective theoretical component target. 
Because the correlation measures the strength of a linear relationship, we call the terms $Cor(p_{x},p_{x}\mu_{x})$ and $Cor(\mu_{x},p_{x}\mu_{x})$ \textit{linear relation factors} in the following. 

Next, we extend our analysis to the case, where both component models are combined to make a prediction for $p_{x}\mu_{x}$. More specifically, we analyse the two component model $\hat{p}(x)\hat{\mu}(x)$. The question to be answered is: If they respectively achieve a performance $Cor(\hat{p}(x),p_{x})$ and $Cor(\hat{\mu}(x),\mu_{x})$ for their theoretical component targets, which performance would their product achieve for the theoretical composite target? In Appendix \ref{app_prod_cor}, we show that this performance is given by
\begin{align}
Cor(\hat{p}(x)\hat{\mu}(x),p_{x}\mu_{x})\approx&w_{\hat{p}\hat{\mu}}Cor(\hat{p}(x),p_{x})Cor(\hat{\mu}(x),\mu_{x})C^{I}+w_{\hat{p}}Cor(\hat{p}(x),p_{x})Cor(p_{x},p_{x}\mu_{x})\nonumber\\
&+w_{\hat{\mu}}Cor(\hat{\mu}(x),\mu_{x})Cor(\mu_{x},p_{x}\mu_{x}), \label{eq_perf_prod}
\end{align} where $w_{\hat{p}\hat{\mu}}$, $w_{\hat{p}}$ and $w_{\hat{\mu}}$ are certain weights, which measure the relative importance of $\hat{p}(x)$, $\hat{\mu}(x)$ and their interaction in the two component model $\hat{p}(x)\hat{\mu}(x)$. For now, we consider these weights as something fixed. In the next section, we will further analyze them and the corresponding problems in the two component model approach. Examining equation \eqref{eq_perf_prod}, we can see that both linear relation factors multiplied by the performance of their respective component models enter into the performance of $\hat{p}(x)\hat{\mu}(x)$. However, there is also a third summand, namely
\begin{align*}
w_{\hat{p}\hat{\mu}}Cor(\hat{p}(x),p_{x})Cor(\hat{\mu}(x),\mu_{x})C^{I}.    
\end{align*} Thereby, $C^{I}$ can be seen as the part of variation in $p_{x}\mu_{x}$, which is due to the interaction of $p_{x}$ and $\mu_{x}$ (see Appendix \ref{app_prod_cor} for details). It is important that $C^{I}$ is multiplied by both component models' respective correlations with their theoretical composite targets. Unless both $Cor(\hat{p}(x),p_{x})$ and $Cor(\hat{\mu}(x),\mu_{x})$ are close to 1, $C^{I}$ gets far less weight in equation \eqref{eq_perf_prod} than $Cor(p_{x},p_{x}\mu_{x})$ and $Cor(\mu_{x},p_{x}\mu_{x})$. This has the following implication: If one or both of the linear relation factors are high and the signal-to-noise ratio of the corresponding target(s) is high, the two component model $\hat{p}(x)\hat{\mu}(x)$ should perform well with the same argumentation as for the individual component models. But, if none of the linear relation factors is high, the two component model could only achieve a high performance, if both $Cor(\hat{p}(x),p_{x})$ and $Cor(\hat{\mu}(x),\mu_{x})$ are very high. In contrast, a single model achieves (per definition) a good performance if $Cor(\hat{\zeta}(x),p_{x}\mu_{x})$ is high. So, only if both $SNR_{c}$ and $SNR_{a}$ are much higher than $SNR_{y}$, would we expect $\hat{p}(x)\hat{\mu}(x)$ to perform better than $\hat{\zeta}(x)$. 

\begin{figure}
\centering
\begin{minipage}{.48\textwidth}
  \centering
  \includegraphics[width=0.93\textwidth]{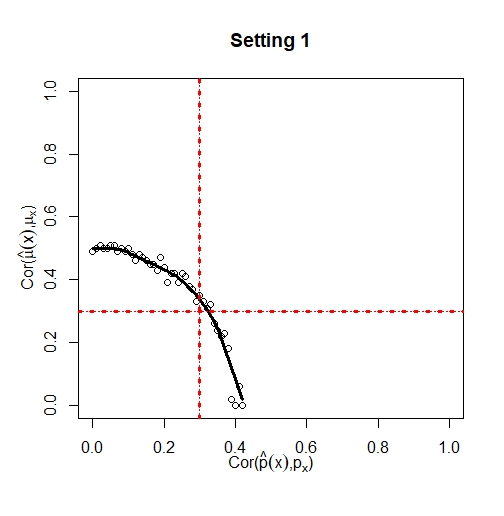}
\end{minipage}%
\hfill
\begin{minipage}{.48\textwidth}
  \centering
  \includegraphics[width=0.93\textwidth]{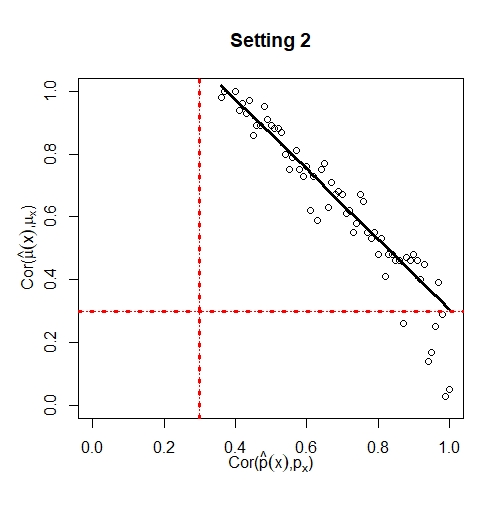}
  
\end{minipage}
\captionof{figure}{The plots show, for the simulation settings in section \ref{sec_comp_models}, how much correlation which their respective component targets the models $\hat{p}(x)$ and $\hat{\mu}(x)$ need to achieve, such that $Cor(\hat{p}(x)\hat{\mu}(x), p_{x}\mu_{x})=0.3$.}
  \label{fig:sim_dist_phatmuhat}
\end{figure}

For an illustration of these rather complicated theoretical properties, we refer to figure \ref{fig:sim_dist_phatmuhat}. This figure presents the results of two simulation scenarios. In the first scenario (left plot), the theoretical component targets have a strong linear relationship to the theoretical composite target. In this case, we can see, that the two component models $\hat{p}(x)$ and $\hat{\mu}(x)$ do not need to achieve a high correlation with their respective theoretical component targets in order for the product to achieve a performance of $Cor(\hat{p}(x)\hat{\mu}(x),p_{x}\mu_{x})=0.3$. In the second scenario (right plot), both the theoretical component targets have a weak linear relationship to the composite target. In this case, we can see that $\hat{p}(x)$ and $\hat{\mu}(x)$ both need a very high performance respectively in order for their product to achieve a correlation of $Cor(\hat{p}(x)\hat{\mu}(x),p_{x}\mu_{x})=0.3$.    

So, to sum up, building predictive models for the component targets $a$ and $c$ and combining their predictions in a prediction for $ac$ makes sense if
\begin{itemize}
    \item[(a)] The signal-to-noise ratio of at least one of the component targets $a$ and $c$ is lower than the signal-to-noise ratio of the composite target $ac$.
    \item[(b)] One or both of the theoretical component targets have a strong linear relationship to the theoretical composite target.
\end{itemize}
If these conditions are not fulfilled, the single model approach can be expected to work better.

\subsection{The scaling and location problem}
In the last section, we derived conditions under which it makes sense to build predictive models for the component targets $a$ and $c$, when the goal is to predict the composite target $ac$. We derived these conditions from equation \eqref{eq_perf_prod}. However, when we did this we ignored one aspect in the equation, namely the weights $w_{\hat{p}\hat{\mu}}$, $w_{\hat{p}}$ and $w_{\hat{\mu}}$. A potential problem which could arise in the two component model approach is that these weights are sub-optimal. Consider equation \eqref{eq_perf_prod} and the exemplary case that $Cor(\hat{p}(x),p_{x})Cor(p_{x},p_{x}\mu_{x})$ is very high. If the weight $w_{\hat{p}}$ would be close to zero, the great performance of $\hat{p}(x)$ would barely affect the performance of the two component model $\hat{p}(x)\hat{\mu}(x)$. Accordingly, the performance of the two component model $\hat{p}(x)\hat{\mu}(x)$ could be far worse than the performance of the component model $\hat{p}(x)$. Next, we show that this can happen in practice and explain why.    

First, we need to examine what the weights $w_{\hat{p}\hat{\mu}}$, $w_{\hat{p}}$ and $w_{\hat{\mu}}$ actually measure. As shown in Appendix \ref{app_weight_comp_models}, they correspond to $CV_{\hat{p}}$, $CV_{\hat{\mu}}$ and $CV_{\hat{p}}\cdot CV_{\hat{\mu}}$ times a certain factor. Here, $CV$ denotes the coefficient of variation (standard deviation divided by the expected value). So, for example, it holds $\frac{w_{\hat{p}}}{w_{\hat{\mu}}}=\frac{CV_{\hat{p}}}{CV_{\hat{\mu}}}$. So, if the model $\hat{p}(x)$ predicts $p_{x}\mu_{x}$ better than the model $\hat{\mu}(x)$ but the coefficient of variation of $\hat{p}(x)$ is lower than the coefficient of variation for $\hat{\mu}(x)$, this will lead to a sub-optimal weighting.

In practice, $\hat{\mu}(x)$ is optimized to predict the target $a$ and $\hat{p}(x)$ is optimized to predict the target $c$. When we apply a simple product of the form $\hat{p}(x)\hat{\mu}(x)$, there are no strong theoretical arguments to assume that the coefficients of variation of these models correspond to their respective performance for predicting $p_{x}\mu_{x}$.   

\section{Applications on real-world data sets}
In this section, we describe our analysis of three real-world applications of predictive modeling for limited dependent variables. We will compare the performance of the single model approach and the two component model approach. Furthermore, we will explain why one of the methods outperformed the other, by applying the results of our theory. 

\subsection{Data sets}
The first data set is from a charity organization, aiming to identify, based on historical data, which persons to send a donation request mail, in order to receive as much donation as possible. The target variable in this data set is \textit{donation amount}, which is zero in 94,9\% of the observations. The data set was retrieved from \citet{kdd}. We only used the training sample, which consists of 95,412 observations. We refer to this data set as "charity" in the following.

The second data set is not publicly available. We obtained it from an online shop. The target variable is \textit{amount spend}, which is zero in 86.7\% of all observations. The data set contains 144,630 observations. We refer to this data set as "private" in the following.

The third data set is also from an online shop. The target variable is again \textit{amount spend}, which is zero in 99.1\% of all cases. We retrieved the data from \citet{hillstrom}. The data set contains 64,000 observations. We refer to this data set as "Hillstrom" in the following.

\subsection{Predictive models}
Each data set, we split into a training set containing 80\% of the observations and a test set containing the remaining 20\% observations. On the training set, we built three models $\hat{\zeta}(x)$ using the observations $(y_{i},x_{i})$, $\hat{p}(x)$ using the observations $(c_{i},x_{i})$ and $\hat{\mu}(x)$ using the observations $(a_{i},x_{i})$, for which $c_{i}=1$. We always applied random forest, using the ranger package \citep{ranger}.

\subsection{Summary statistics}
The first summary statistic we applied to each data set was $Cor(ac,c)$. This correlation measures how strong $ac$ linearly depends on $c$. If $Cor(ac,c)\approx 1$, we know, that $ac$ can not contain much considerably more information than $c$ as a prediction target. So, a correlation significantly smaller than 1, tells us that there could be some valuable information in $a$, which we might want to leverage.

The next summary statistics, we consider, are $Cor(\hat{p}(x),c)$ and $Cor(\hat{\mu}(x),a)$ (calculated on the observations with $c=1$). These correlations measure the performance of the component models with regard to the observable targets. What is most interesting about them is that they are indicators for the signal-to-noise ratios $SNR_{c}$ and $SNR_{a}$. As outlined in section \ref{sec_sig_noise}, low correlations can be seen as indicators of a low signal-to-noise ratio and high correlations can be seen as indicators of a high signal-to-noise ratio.

Finally, we examine the performance measures $Cor(\hat{p}(x),ac)$, $Cor(\hat{p}(x)\hat{\mu}(x),ac)$ and $Cor(\hat{\zeta}(x),ac)$. These correlations measure the performance of the respective models. With the help of the other summary statistics, we can explain why either $\hat{\zeta}(x)$, $\hat{p}(x)$ or $\hat{p}(x)\hat{\mu}(x)$ performed best.

In a separate analysis, we examined the scaling and location problem in the component model approach. Therefore, we calculated for different \textit{adjustment factors} $s_{ad}$, the performance $Cor(\hat{p}(x)\hat{\mu}_{s}(x))$, where $\hat{\mu}_{s}(x))$ is the model $\hat{\mu}(x))$, only with an adjusted coefficient of variation, in the following way
\begin{align*}
\hat{\mu}_{s}(x))=(\hat{\mu}(x))-E[\hat{\mu}(x))])\cdot s_{ad}+ E[\hat{\mu}(x))].   
\end{align*} We analyzed, for which adjustment factor the highest performance was achieved.

\section{Results}
The results are provided in table \ref{tab_cor}. We can see, that on each real-world data set either $\hat{p}(x)$ or $\hat{p}(x)\hat{\mu}(x)$ performed best, but never $\hat{\zeta}(x)$. On the "charity" data set, $\hat{p}(x)\hat{\mu}(x)$ outperformed the other approaches by far. This is in line with the results of \citet{zadrozny2001learning}, who also found the two component model to outperform the single model on this data set. The result can easily be explained: We can see that $Cor(ac,c)$ is significantly lower than 1 and $Cor(\hat{\mu}(x),a)$ is quite high. This implies that $a$ has a significant impact on the distribution of $ac$ and that the signal-to-noise ratio of $a$ is high. Seeing that $Cor(\hat{p}(x),c)$ and $Cor(\hat{\zeta}(x),ac)$ are much lower than $Cor(\hat{\mu}(x),a)$, it is almost certain, that the key to success was the increase in the signal-to-noise ratio of $a$, when the target $ac$ was separated into its components.

For the "private" data set and the "Hillstrom" data set, $\hat{p}(x)$ achieved the highest performance. Again, it seems plausible that the increase in the signal-to-noise ratio by separation of $ac$ in its components was the key to success. Only, that, in contrast to the "charity" data, here the target $c$ seems to be more valuable than $a$ (as can be seen by $Cor(\hat{p}(x),c)>Cor(\hat{\mu}(x),a)$ and $Cor(\hat{p}(x),c)>Cor(\hat{\zeta}(x),ac)$). On the "Hillstrom" data, $\hat{\mu}(x)$ even has a negative correlation with $a$. So, it is very plausible that $\hat{p}(x)\hat{\mu}(x)$ performs worse than $\hat{p}(x)$. But what about the "private" data set. There $\hat{\mu}(x)$ has a positive correlation with $a$. Why, does this not make $\hat{p}(x)\hat{\mu}(x)$ perform better than $\hat{p}(x)$? The reason lies in the scaling and location problem.

\begin{figure}
\centering
\begin{minipage}{.48\textwidth}
  \centering
  \includegraphics[width=0.93\textwidth]{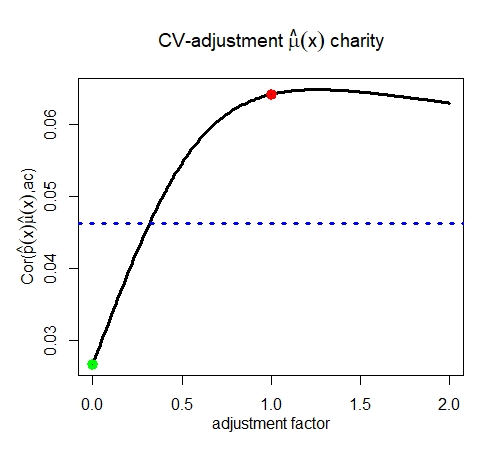}
\end{minipage}%
\hfill
\begin{minipage}{.48\textwidth}
  \centering
  \includegraphics[width=0.93\textwidth]{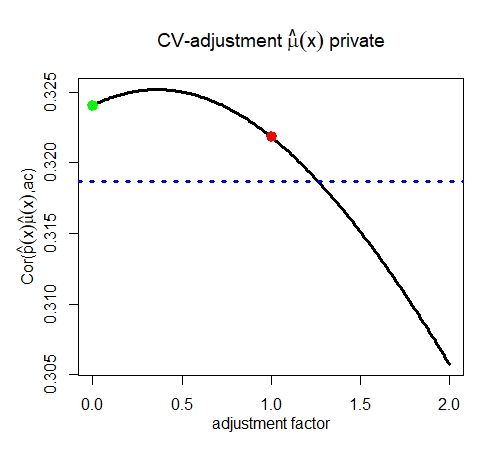}
  
\end{minipage}

\begin{minipage}{.48\textwidth}
  \centering
  \includegraphics[width=0.93\textwidth]{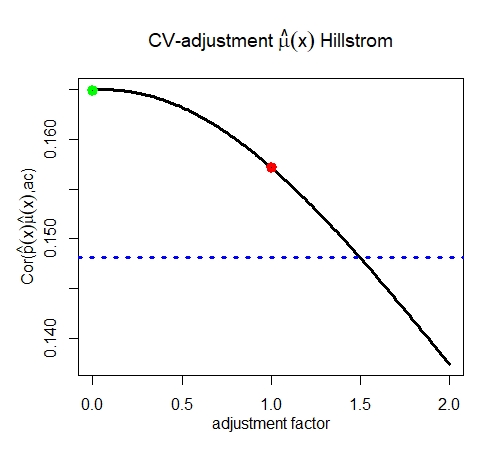}
\end{minipage}%
\hfill
\begin{minipage}{.48\textwidth}
  \centering
  \includegraphics[width=0.93\textwidth]{CV_ad_Hillstrom.jpeg}
  
\end{minipage}
\captionof{figure}{The plots show the correlation of $\hat{p}(x)\hat{\mu}(x)$ and $ac$, for different CV-adjustment factors of $\hat{\mu}(x)$. The green and red dots represent the results for adjustment factors $0$ and $1$, which corresponds to $\hat{p}(x)$ and $\hat{p}(x)\hat{\mu}(x)$ without adjustment. The blue line is the correlation between $ac$ and the single model $\hat{\zeta}(x)$.}
  \label{fig:CV_ad}
\end{figure}

The scaling and location problem is analyzed in figure \ref{fig:CV_ad}. We can see, that for none of the data sets, the original scaling of $\hat{p}(x)\hat{\mu}(x)$ was ideal, because none of the curves has its maximum at $s_{ad}=1$. But it needs being said that on the "charity" data set, the original scaling is close to optimal. On the "Hillstrom" data set the optimal scaling is at $s_{ad}=0$, which would correspond to $\hat{p}(x)$ without any influence of $\hat{\mu}(x)$. The "private" data set represents the most interesting case. The optimal scaling would be at $s_{ad}\approx 0.5$. With this scaling, the two component model approach would perform better than $\hat{p}(x)$. Without this downscaling of the importance of $\hat{\mu}(x)$, the two component model approach performs worse than $\hat{p}(x)$.   

\begin{table*}[hbt]
\begin{center}
\caption{Considered simulation settings}\label{tab_cor}
\begin{threeparttable}

\begin{tabular*}{\textwidth}{p{0.1\textwidth}p{0.1\textwidth}p{0.1\textwidth}p{0.15\textwidth}p{0.1\textwidth}p{0.12\textwidth}p{0.12\textwidth}}
    \toprule
    \textbf{data} & $\bm{Cor(ac,c)}$ & $\bm{Cor(\hat{p},c)}$ & $\bm{Cor(\hat{\mu},a)|_{c=1}}$ & $\bm{Cor(\hat{p},ac)}$ & $\bm{Cor(\hat{p}\hat{\mu},ac)}$ & $\bm{Cor(\hat{\zeta},ac)}$\\
    \hline\hline
    'charity' & 0.77  & 0.068 & 0.765 & 0.027 & 0.064 & 0.046 \\
    'private' &  0.774 & 0.421 & 0.175 & 0.324 & 0.322 & 0.319 \\
    'Hillstrom' & 0.732 & 0.200 & -0.018 & 0.165 & 0.157 & 0.148 \\
    \hline

\end{tabular*}

\begin{tablenotes}[para,flushleft]
   \small
   The table shows the results for the three real-world data sets.    
   \end{tablenotes}

\end{threeparttable}
\end{center}
\end{table*}

\appendix
\section{Relationship between $c$ and $a$}\label{app_sig_noise}
In the following, we derive an expression for the variance of $ac$. Before we do this, we note that the expected value of $ac$ can be written as
\begin{align*}
E[ac]&=P[c=1]E[a|c=1]\\
&=E[p_{x}]E[\mu_{x}|c=1].
\end{align*} Now, we come to the variance of $ac$.
\begin{align}
Var[ac]&=E[(ac-E[ac])^2]\nonumber\\
&=E[(ac-E[p_{x}]E[\mu_{x}|c=1])^2]\nonumber\\
&=P[c=1]E[(ac-E[p_{x}]E[\mu_{x}|c=1])^2|c=1]+P[c=0]E[(ac-E[p_{x}]E[\mu_{x}|c=1])^2|c=0]\nonumber\\
&=P[c=1]E[(a-E[p_{x}]E[\mu_{x}|c=1])^2|c=1]+P[c=0]E[p_{x}]^2E[\mu_{x}|c=1]^2\nonumber\\
&=E[p_{x}]E[(a-E[p_{x}]E[\mu_{x}|c=1])^2|c=1]+(1-E[p_{x}])E[p_{x}]^2E[\mu_{x}|c=1]^2\nonumber\\
&=E[p_{x}]E[a^2|c=1]-2E[p_{x}]^2E[\mu_{x}|c=1]^2+E[p_{x}]^3E[\mu_{x}|c=1]^2+(1-E[p_{x}])E[p_{x}]^2E[\mu_{x}|c=1]^2\nonumber\\
&=E[p_{x}]E[a^2|c=1]-E[p_{x}]^2E[\mu_{x}|c=1]^2\nonumber\\
&=E[p_{x}]E[a^2|c=1]-E[p_{x}]E[\mu_{x}|c=1]^2+(E[p_{x}]-E[p_{x}]^2)E[\mu_{x}|c=1]^2\nonumber\\
&=E[p_{x}]Var[a|c=1]+Var[c]E[\mu_{x}|c=1]^2.\label{eq_var_ac_dec}
\end{align} With this expression for the variance, we can derive an expression for the correlation between $c$ and $ac$.
\begin{align*}
Cov(ac,c)&=E[ac^2]-E[ac]E[c]\\
&=E[ac]-E[ac]E[c]\\
&=E[p_{x}]E[\mu_{x}|c=1]-E[p_{x}]E[\mu_{x}|c=1]E[p_{x}]\\
&=Var[c]E[\mu_{x}|c=1]
\end{align*} Accordingly,
\begin{align*}
Cor(ac,c)^2&=\frac{Cov(ac,c)^2}{Var[ac]Var[c]}\\
&=\frac{Var[c]E[\mu_{x}|c=1]^2}{Var[ac]}\\
&=\frac{Var[c]E[\mu_{x}|c=1]^2}{E[p_{x}]Var[a|c=1]+Var[c]E[\mu_{x}|c=1]^2}\\
&=\frac{1}{1+\frac{E[p_{x}]}{Var[c]}\frac{Var[a|c=1]}{E[\mu_{x}|c=1]^2}}\\
&=\frac{1}{1+\frac{1}{1-E[p_{x}]}\frac{Var[a|c=1]}{E[\mu_{x}|c=1]^2}}\\
&=\frac{1}{1+\frac{CV_{a}^2}{1-E[p_{x}]}}.
\end{align*} 

Now, we derive the relationship between $E[\mu_{x}]$ and $E[\mu_{x}|c=1]$. It holds
\begin{align*}
E[p_{x}\mu_{x}]&=E[P[c=1|x]\cdot\mu_{x}]\\
&=E[p_{x}]E[\mu_{x}|c=1].    
\end{align*} With $E[p_{x}\mu_{x}]=Cov(p_{x},\mu_{x})+E[p_{x}]\cdot E[\mu_{x}]$, we obtain
\begin{align}
E[\mu_{x}]=E[\mu_{x}|c=1]-\frac{Cov(p_{x},\mu_{x})}{E[p_{x}]}.\label{eq_mu_cond}    
\end{align}

\section{Derivations of the signal-to-noise ratio}\label{app_sig_noise}
In Appendix \ref{app_sig_noise}, we show that the variance of $ac$ is given by
\begin{align*}
Var[ac]=E[p_{x}]Var[a|c=1]+E[\mu_{x}|c=1]^2Var[c].    
\end{align*} Now, we will find out how much of this is signal and how much of this is noise. To this end, we need to derive the following equation for the noise $Var[\varepsilon]$ in $ac$:
\begin{align*}
Var[\varepsilon]&=E[(p_{x}\varepsilon_{\mu}+\mu_{x}\varepsilon_{p}+\varepsilon_{p}\varepsilon_{\mu})^2]\\
&=E[(c\cdot\varepsilon_{\mu}+\mu_{x}\cdot\varepsilon_{p})^2]\\
&=E[c\cdot\varepsilon_{\mu}^2]+E[\mu_{x}^2\cdot\varepsilon_{p}^2]\\
&=E[p_{x}]Var[\varepsilon_{\mu}^2|c=1]+E[\mu_{x}^2\cdot\varepsilon_{p}^2].
\end{align*} The first summand does not require additional analysis. The second summand needs some further examination. It holds
\begin{align*}
E[\mu_{x}^2\cdot\varepsilon_{p}^2]&=E[\mu_{x}^2]\cdot Var[\varepsilon_{p}]+Cov(\mu_{x}^2, \varepsilon_{p}^2)\\
&=E[\mu_{x}]^2\cdot Var[\varepsilon_{p}]+Var[\mu_{x}]\cdot Var[\varepsilon_{p}]+Cov(\mu_{x}^2, \varepsilon_{p}^2) \text{ with equation \eqref{eq_mu_cond}}\\
&=\left(E[\mu_{x}|c=1]-\frac{Cov(p_{x},\mu_{x})}{E[p_{x}]}\right)^2\cdot Var[\varepsilon_{p}]+Var[\mu_{x}]\cdot Var[\varepsilon_{p}]+Cov(\mu_{x}^2, \varepsilon_{p}^2).
\end{align*} Apparently, this noise term depends on the statistical relationship between $a$ and $c$ (expressed by $Cov(p_{x},\mu_{x})$ and $Cov(\mu_{x}^2, \varepsilon_{p}^2)$). 

For the sake of simplicity, we first consider the case that $a$ and $c$ are independent. In this case, the expression for $E[\mu_{x}^2\cdot\varepsilon_{p}^2]$ simplifies to
\begin{align*}
E[\mu_{x}^2\cdot\varepsilon_{p}^2]=E[\mu_{x}|c=1]^2\cdot Var[\varepsilon_{p}].
\end{align*} So signal and noise of $ac$ become accordingly
\begin{align*}
Var[p_{x}\mu_{x}]&=E[p_{x}]Var[\mu_{x}|c=1]+E[\mu_{x}|c=1]^2Var[p_{x}]-Var[\mu_{x}]\cdot Var[\varepsilon_{p}]\\
Var[\varepsilon]&=E[p_{x}]Var[\varepsilon_{\mu}|c=1]+E[\mu_{x}|c=1]^2Var[\varepsilon_{p}]+Var[\mu_{x}]\cdot Var[\varepsilon_{p}].
\end{align*} As $Var[\mu_{x}]\cdot Var[\varepsilon_{p}]$ always decreases the signal and increases the noise of $y=ac$, it holds for the signal-to-noise ratio of $y$ 
\begin{align*}
SNR_{y}&<\alpha\cdot SNR_{a}+(1-\alpha)SNR_{c} \text{ with}\\
\alpha&:=\frac{E[p_{x}]Var[\varepsilon_{\mu}|c=1]}{E[p_{x}]Var[\varepsilon_{\mu}|c=1]+E[\mu_{x}|c=1]^2Var[\varepsilon_{p}]}.
\end{align*} So, $SNR_{y}$ is less than a weighted sum of $SNR_{a}$ and $SNR_{c}$.

Now, we examine the case when $a$ and $c$ are not independent. In this case, signal and noise of $ac$ become
\begin{align*}
Var[p_{x}\mu_{x}]&=E[p_{x}]Var[\mu_{x}|c=1]+E[\mu_{x}|c=1]^2Var[p_{x}]-D_{\varepsilon}\\
Var[\varepsilon]&=E[p_{x}]Var[\varepsilon_{\mu}|c=1]+E[\mu_{x}|c=1]^2Var[\varepsilon_{p}]+D_{\varepsilon},    
\end{align*} where the dependent component $D_{\varepsilon}$ is given by
\begin{align*}
 D_{\varepsilon}:=Var[\mu_{x}]\cdot Var[\varepsilon_{p}]+\left(\frac{Cov(p_{x},\mu_{x})}{E[p_{x}]}-2\cdot E[\mu_{x}|c=1]\right)\cdot\frac{Cov(p_{x},\mu_{x})}{E[p_{x}]}\cdot Var[\varepsilon_{p}]+Cov(\mu_{x}^2,\varepsilon_{p}^2).    
\end{align*} Only, if the statistical dependence between $a$ and $c$ is such, that 
\begin{align*}
-\left(\frac{Cov(p_{x},\mu_{x})}{E[p_{x}]}-2\cdot E[\mu_{x}|c=1]\right)\cdot\frac{Cov(p_{x},\mu_{x})}{E[p_{x}]}\cdot Var[\varepsilon_{p}]+Cov(\mu_{x}^2,\varepsilon_{p}^2)\geq Var[\mu_{x}]\cdot Var[\varepsilon_{p}],
\end{align*} the signal-to-noise ratio $SNR_{y}$ would be higher than the weighted sum $\alpha\cdot SNR_{a}+(1-\alpha)SNR_{c}$. Else, it stays lower than this weighted sum. In any case, the signal-to-noise ratio of $y$ can be written as
\begin{align*}
SNR_{y}=\alpha\cdot SNR_{a}+(1-\alpha)SNR_{c}+\Psi,
\end{align*} where, depending on the relationship between $a$ and $c$, $\Psi$ could be positive or negative. 

\section{Relationship to sample selection bias}\label{app_sel_bias}
The potential problem of sample selection bias concern the fact, that model $\hat{\mu}(x)$, in a two component model, is trained only on the sub-sample of observations with $c=1$ and applied on all observations (where $c$ could either be 1 or 0).\citep{zadrozny2001learning} In the following, we analyze in what way training $\hat{\mu}(x)$ on the sub-sample might impact the model when making predictions using the two component model $\hat{p}(x)\hat{\mu}(x)$ for $y$.

We know from section \ref{sec_def_target}, that the conditional expected value of $y$ is given as $E[y|x]=p_{x}\mu_{x}$. We further know, that $\hat{p}(x)$ is trained on the whole sample for the target $c$. Hence, it is correctly trained to yield predictions for $E[c|x]=p_{x}$. In contrast, $\hat{\mu}(x)$ is trained only on observations with $c=1$. Hence, it is correctly trained to yield predictions for $E[a|x,c=1]=\mu_{x}$. Accordingly, the product of $\hat{p}(x)\mu_{x}$ is a plausible predictive models for $p_{x}\mu_{x}$. So, where is the problem of sample selection bias?

It is true, that $\hat{\mu}(x)$ is trained on the sub-sample with $c=1$ and applied to the whole sample. However, this does not render its predictions biased. For each point $x$, $\hat{\mu}(x)$ needs to make a prediction for $E[a|x,c=1]$ and this is exactly, what $\hat{\mu}(x)$ is trained for. The only important aspect of the sample selection is that observations with features $x$, for which $P[c=1|x]$ is low, will seldom be in the training sample for $\hat{\mu}(x)$. For such observations the accuracy will generally be lower than for observations with features having a high $P[c=1|x]$. But this is not a model bias for which correction is needed.

The application for which Heckman suggested his sample selection bias correction is different from our predictive modeling application.\citep{heckman1979sample}. He described applications where one is interested in \textit{estimating average effects of some covariates} for the whole population, in situations where the outcome $y$ is only observed in a sub-sample ($c=1$). For such an application, the sample selection is indeed a problem, because the average effects measured on a sub-population can differ from the average effects of covariates for the whole population. Accordingly, it is plausible to apply a strategy to correct for the potential bias. However, in our predictive modeling application, model $\hat{\mu}(x)$ should predict the expected outcome for the sub-sample it is trained on. So, there is no need for sample selection correction. 

The only thing to consider would be weighting the observations which $\hat{\mu}(x)$ is trained on (observations with $c=1$) according to their likelihood to appear in the whole sample (observations with $c=1$ and $c=0$). This could be done by using the estimated inverse probability $1/P[c|x]$. In this way, $\hat{\mu}(x)$ would still be trained for the right theoretical parameter $\mu_{x}$ and possibly its performance on the whole for predicting $\mu_{x}$ would increase. However, we decided not to examine this weighting approach, to keep a clear scope of the research project.   

\section{Performance of the component target models}

\subsection{Performance of a model for the target $c$}\label{app_class_model}
We start by deriving an expression for $Cor(p_{x}\mu_{x},\hat{p}(x))$. Using known results from linear regression modeling, we know that we can write the predictions of a $\hat{p}(x)$ as 
\begin{align*}
\hat{p}(x)&=\beta + \alpha p_{x}+\varepsilon_{1} \text{ with}\\
\alpha&=\frac{Cov(\hat{p}(x),p_{x})}{Var[p_{x}]}\\
Cov(p_{x},\varepsilon_{1})&=0.
\end{align*} Then, we can write the covariance between $\hat{p}(x)$ and the target $p_{x}\mu_{x}$ as
\begin{align*}
Cov(\hat{p}(x),p_{x}\mu_{x})&=Cov(\beta + \alpha p_{x}+\varepsilon_{1},p_{x}\mu_{x})\\
&=\alpha Cov(p_{x},p_{x}\mu_{x})+Cov(\varepsilon_{1},p_{x}\mu_{x})\\
&=\frac{Cov(\hat{p}(x),p_{x})Cov(p_{x},p_{x}\mu_{x})}{Var[p_{x}]}+Cov(\varepsilon_{1},p_{x}\mu_{x}).
\end{align*} So, it holds for the correlation
\begin{align}
Cor(\hat{p}(x),p_{x}\mu_{x})&=Cor(\hat{p}(x),p_{x})Cor(p_{x},p_{x}\mu_{x})+Cor(\varepsilon_{1},p_{x}\mu_{x})\sqrt{\frac{Var[\varepsilon_{1}]}{Var[\hat{p}(x)]}}\nonumber\\ 
&=Cor(\hat{p}(x),p_{x})Cor(p_{x},p_{x}\mu_{x})+Cor(\varepsilon_{1},p_{x}\mu_{x})\sqrt{1-Cor(\hat{p}(x),p_{x})^2}.\label{eq_cor_p}
\end{align} Here, the last equation follows from the definition of $\varepsilon_{1}$ and the fact that $Cov(\varepsilon_{1},p_{x})=0$. Note, that the rest term $\varepsilon_{1}$ is uncorrelated with the the signal $p_{x}$ in the component target $c$, used to train $\hat{p}(x)$. Hence, $\varepsilon_{1}$ can be interpreted as part of the variation in $\hat{p}(x)$, which is due to the noise $\varepsilon_{p}$ in the training data for $\hat{p}(x)$. As this noise in the training data is not correlated with $\mu_{x}$, there is no reason to assume a relevant correlation $Cor(\varepsilon_{1},p_{x}\mu_{x})$. Accordingly, it holds approximately
\begin{align*}
Cor(\hat{p}(x),p_{x}\mu_{x})\approx Cor(\hat{p}(x),p_{x})Cor(p_{x},p_{x}\mu_{x}).  
\end{align*}

In the same way, we can derive
\begin{align*}
Cor(\hat{\mu}(x),p_{x}\mu_{x})&=Cor(\hat{\mu}(x),\mu_{x})Cor(\mu_{x},p_{x}\mu_{x})+Cor(\varepsilon_{2},p_{x}\mu_{x})\sqrt{1-Cor(\hat{\mu}(x),\mu_{x})^2}\\
&\approx Cor(\hat{\mu}(x),\mu_{x})Cor(\mu_{x},p_{x}\mu_{x}),    
\end{align*} where $\varepsilon_{2}$ is the rest term in the linear projection from $\hat{\mu}(x)$ on $\mu_{x}$.

\subsection{Performance of the two component model $\hat{p}(x)\hat{\mu}(x)$}\label{app_prod_cor}
Now, we derive an expression for $Cor(p_{x}\mu_{x},\hat{p}(x)\hat{\mu}(x))$. We can write the component models as
\begin{align*}
\hat{p}(x)&=\beta_{1}+\alpha_{1}p_{x}+\varepsilon_{1} \text{ with } Cov(\varepsilon_{1},p_{x})=0\\
\hat{\mu}(x)&=\beta_{2}+\alpha_{2}\mu_{x}+\varepsilon_{2} \text{ with } Cov(\varepsilon_{2},\mu_{x})=0\\
\beta_{1}&=E[\hat{p}(x)]-\alpha_{1}E[p_{x}]\\
\beta_{2}&=E[\hat{\mu}(x)]-\alpha_{2}E[\mu_{x}]\\
\alpha_{1}&=\frac{Cov(\hat{p}(x),p_{x})}{Var[p_{x}]}\\
\alpha_{2}&=\frac{Cov(\hat{\mu}(x),\mu_{x})}{Var[\mu_{x}]}.
\end{align*} So it holds
\begin{align*}
Cov(p_{x}\mu_{x},\hat{p}(x)\hat{\mu}(x))=&\alpha_{1}\alpha_{2}Var[p_{x}\mu_{x}]+\alpha_{1}\beta_{2}Cov(p_{x}\mu_{x},p_{x})+\alpha_{2}\beta_{1}Cov(p_{x}\mu_{x},\mu_{x})+Cov(p_{x}\mu_{x},\varepsilon_{1}\hat{\mu}(x))\\
&+Cov(p_{x}\mu_{x},\varepsilon_{2}\hat{p}(x)). 
\end{align*} We define
\begin{align*}
C_{\hat{p}\hat{\mu}}&=\frac{\alpha_{1}\alpha_{2}Var[p_{x}\mu_{x}]+\alpha_{1}\beta_{2}Cov(p_{x}\mu_{x},p_{x})+\alpha_{2}\beta_{1}Cov(p_{x}\mu_{x},\mu_{x})}{\sqrt{Var[p_{x}\mu_{x}]Var[\hat{p}(x)\hat{\mu}(x)]}}\\
R_{\hat{p}\hat{\mu}}&=\frac{Cov(p_{x}\mu_{x},\varepsilon_{1}\hat{\mu}(x))+Cov(p_{x}\mu_{x},\varepsilon_{2}\hat{p}(x))}{\sqrt{Var[p_{x}\mu_{x}]Var[\hat{p}(x)\hat{\mu}(x)]}} \text{ such that}\\
Cor(p_{x}\mu_{x},\hat{p}(x)\hat{\mu}(x))&=C_{\hat{p}\hat{\mu}}+R_{\hat{p}\hat{\mu}}.
\end{align*} With a similar argumentation than for the component target models we can derive that $R_{\hat{p}\hat{\mu}}\approx 0$: $\varepsilon_{1}$ and $\varepsilon_{2}$ describe variation in $\hat{p}(x)$ and $\hat{\mu}(x)$, which has zero covariance with their respective theoretical targets. Only by chance would the covariance with $p_{x}\mu_{x}$ be different from zero. The same holds for $Cov(\varepsilon_{1},\hat{\mu}(x))$ and $Cov(\varepsilon_{2},\hat{p}(x))$. Putting this together, we can expect $R_{\hat{p}\hat{\mu}}\approx 0$.   

So, it holds $Cor(p_{x}\mu_{x},\hat{p}(x)\hat{\mu}(x))\approx C_{\hat{p}\hat{\mu}}$. Next, we further examine $C_{\hat{p}\hat{\mu}}$. It holds 
\begin{align*}
C_{\hat{p}\hat{\mu}}=&Cor(\hat{p}(x),p_{x})Cor(\hat{\mu}(x),\mu_{x})\sqrt{\frac{Var[\hat{p}(x)]}{Var[p_{x}]}}\sqrt{\frac{Var[\hat{\mu}(x)]}{Var[\mu_{x}]}}\sqrt{\frac{Var[p(x)\mu(x)]}{Var[\hat{p}(x)\hat{\mu}(x)]}}\\
&+Cor(\hat{p}(x),p_{x})Cor(p_{x}\mu_{x},p_{x})\left(E[\hat{\mu}(x)]-Cor(\hat{\mu}(x),\mu_{x})\sqrt{\frac{Var[\hat{\mu}(x)]}{Var[\mu_{x}]}}E[\mu_{x}]\right)\sqrt{\frac{Var[\hat{p}(x)]}{Var[\hat{p}(x)\hat{\mu}(x)]}}\\
&+Cor(\hat{\mu}(x),\mu_{x})Cor(p_{x}\mu_{x},\mu_{x})\left(E[\hat{p}(x)]-Cor(\hat{p}(x),p_{x})\sqrt{\frac{Var[\hat{p}(x)]}{Var[p_{x}]}}E[p_{x}]\right)\sqrt{\frac{Var[\hat{\mu}(x)]}{Var[\hat{p}(x)\hat{\mu}(x)]}}\\
=&Cor(\hat{p}(x),p_{x})Cor(\hat{\mu}(x),\mu_{x})\sqrt{\frac{Var[\hat{p}(x)]Var[\hat{\mu}(x)]}{Var[\hat{p}(x)\hat{\mu}(x)]}}\\
&\cdot\left(\sqrt{\frac{Var[p(x)\mu(x)]}{Var[p(x)]Var[\mu(x)]}}
-Cor(p_{x}\mu_{x},p_{x})\frac{E[\mu_{x}]}{\sqrt{Var[\mu_{x}]}}-Cor(p_{x}\mu_{x},\mu_{x})\frac{E[p_{x}]}{\sqrt{Var[p_{x}]}} \right)\\
&+Cor(\hat{p}(x),p_{x})Cor(p_{x}\mu_{x},p_{x})\sqrt{\frac{E[\hat{\mu}(x)]^2Var[\hat{p}(x)]}{Var[\hat{p}(x)\hat{\mu}(x)]}}+Cor(\hat{\mu}(x),\mu_{x})Cor(p_{x}\mu_{x},\mu_{x})\sqrt{\frac{E[\hat{p}(x)]^2Var[\hat{\mu}(x)]}{Var[\hat{p}(x)\hat{\mu}(x)]}}.\\
\end{align*} In the following, we want to interpret the expression for $C_{\hat{p}\hat{\mu}}$. To this end, we first note, that the expression contains three terms expressing variance components of $Var[\hat{p}(x)\hat{\mu}(x)]$. We can interpret these terms as weights for the individual component models in the product $\hat{p}(x)\hat{\mu}(x)$. These terms are
\begin{align*}
w_{\hat{p}\hat{\mu}}:=\sqrt{\frac{Var[\hat{p}(x)]Var[\hat{\mu}(x)]}{Var[\hat{p}(x)\hat{\mu}(x)]}}\\
w_{\hat{p}}:=\sqrt{\frac{E[\hat{\mu}(x)]^2Var[\hat{p}(x)]}{Var[\hat{p}(x)\hat{\mu}(x)]}}\\
w_{\hat{\mu}}:=\sqrt{\frac{E[\hat{p}(x)]^2Var[\hat{\mu}(x)]}{Var[\hat{p}(x)\hat{\mu}(x)]}}.
\end{align*} We further define
\begin{align*}
C^{I}:=\left(\sqrt{\frac{Var[p(x)\mu(x)]}{Var[p(x)]Var[\mu(x)]}}
-Cor(p_{x}\mu_{x},p_{x})\frac{E[\mu_{x}]}{\sqrt{Var[\mu_{x}]}}-Cor(p_{x}\mu_{x},\mu_{x})\frac{E[p_{x}]}{\sqrt{Var[p_{x}]}} \right).
\end{align*} Then we obtain the following expression
\begin{align*}
C_{\hat{p}\hat{\mu}}=&w_{\hat{p}\hat{\mu}}Cor(\hat{p}(x),p_{x})Cor(\hat{\mu}(x),\mu_{x})C^{I}+w_{\hat{p}}Cor(\hat{p}(x),p_{x})Cor(p_{x},p_{x}\mu_{x})\nonumber\\
&+w_{\hat{\mu}}Cor(\hat{\mu}(x),\mu_{x})Cor(\mu_{x},p_{x}\mu_{x}).    
\end{align*}   

\subsection{Weights of the component models}\label{app_weight_comp_models}
In the last appendix section, we have derived weights $w_{\hat{p}\hat{\mu}}$, $w_{\hat{p}}$, $w_{\hat{\mu}}$ in the expression for $C_{\hat{p}\hat{\mu}}\approx Cor(\hat{p}(x)\hat{\mu}(x),p_{x}\mu_{x})$. Here, we want to further examine their meaning. We can write
\begin{align*}
w_{\hat{p}\hat{\mu}}&=\frac{E[\hat{p}(x)]\cdot E[\hat{\mu}(x)]}{\sqrt{Var[\hat{p}(x)\hat{\mu}(x)]}}\cdot\frac{\sqrt{Var[\hat{p}(x)]}}{E[\hat{p}(x)]}\cdot\frac{\sqrt{Var[\hat{\mu}(x)]}}{E[\hat{\mu}(x)]}\\
w_{\hat{p}}&=\frac{E[\hat{p}(x)]\cdot E[\hat{\mu}(x)]}{\sqrt{Var[\hat{p}(x)\hat{\mu}(x)]}}\cdot \frac{\sqrt{Var[\hat{p}(x)]}}{E[\hat{p}(x)]}\\
w_{\hat{\mu}}&=\frac{E[\hat{p}(x)]\cdot E[\hat{\mu}(x)]}{\sqrt{Var[\hat{p}(x)\hat{\mu}(x)]}}\cdot \frac{\sqrt{Var[\hat{\mu}(x)]}}{E[\hat{\mu}(x)]}.
\end{align*} The three weights have the same factor to the left. The right factors correspond to the coefficients of variation (defined as standard deviation divided by the expected value of a random variable) $CV_{\hat{p}}$ and $CV_{\hat{\mu}}$. So, the relative magnitude of the weights to each other correspond to the relative magnitude of $CV_{\hat{p}}\cdot CV_{\hat{\mu}}$ to $CV_{\hat{p}}$ to $CV_{\hat{\mu}}$.

\bibliographystyle{unsrtnat}
\bibliography{template}

\end{document}